\documentclass[aps,preprint,showpacs,superscriptaddress,nofootinbib]{revtex4}
 \usepackage{graphicx,amsmath,amssymb}
\usepackage[usenames]{color}
\usepackage[colorlinks=true, urlcolor=navyblue, linkcolor=navyblue, citecolor=navyblue]{hyperref}
\usepackage{relsize}
\definecolor{navyblue}{rgb}{0,0.08,0.45}
\def\Journal#1#2#3#4{{#1} {{\bf #2},} {#3} {(#4)}}

\def\NPB{{Nucl. Phys.} B }

\def\PRL{ Phys. Rev. Lett.}

\def\PRD{{Phys. Rev.} D}
\def\PRC{{Phys. Rev.} C}

\def\CPC{Chin. Phys. C}
\def\JHEP{{J. High Energy Phys.}}

\def\IJMPA{Int. J. Mod. Phys. A}

\def\be{\begin{equation}}
\def\ee{\end{equation}}
\def\bea{\begin{eqnarray}}
\def\eea{\end{eqnarray}}

\def\babar{\mbox{\slshape B\kern-0.1em{\smaller A}\kern-0.1em
    B\kern-0.1em{\smaller A\kern-0.2em R}}}

\begin{document}

\title{Extraction of neutrino mixing parameters from experiments with multiple identical detectors}
\author{Fu-Guang Cao}
\affiliation{Institute of Fundamental Sciences, Massey University, Private Bag 11 222, Palmerston North, New Zealand}
\author{William S. Marks}
\affiliation{Institute of Fundamental Sciences, Massey University, Private Bag 11 222, Palmerston North, New Zealand}

\date{\today}

\begin{abstract}
The statistical method used in the analyses of measurements of neutrino oscillation mixing angle $\theta_{13}$ by 
the Daya Bay Collaboration  \cite{DB2012} and RENO Collaboration \cite{RENO2012} is based on variational minimization of a $\chi^2$ function defined
in terms of quantities of interest and pull factors which are introduced to deal with effects of
systematic uncertainties. For both experiments, the number of parameters that need to be determined is great than the number of available data points
(20 vs 6 for the Daya Bay and 12 vs 2 for the RENO). 
While the results for the mixing angle and the normalization factor were reported, results for the other parameters (pull factors) were omitted
in their publications \cite{DB2012,RENO2012}.
There exist multiple sets of parameters from the minimization of the $\chi^2$ function. 
We investigate the sensitivity of the extracted mixing angle on this non-uniqueness of minimization results for the Daya Bay data using two methods of minimization.
We report results for all parameters, including those of physics interest and pull factors.
The obtained results for the mixing angle and the normalization factor are in agreement with those reported by the Daya Bay Collaboration. 
Furthermore, we present plots of confidence level contours in the space of the mixing angle and normalization factor.
We also present results from fittings using a reduced $\chi^2$ function with fewer parameters than the one employed by the Daya Bay Collaboration.
\end{abstract}

\maketitle

\section{Introduction}

The parameters in the Pontecorvo-Maki-Nakagawa-Sakata (PMNS) matrix \cite{PMNS}, the currently accepted framework describing neutrino oscillations,
can only be determined experimentally.
While the measurements of mixing angles $\theta_{12}$ and $\theta_{23}$ are achieved with solar neutrinos and
atmospheric neutrinos (with the help of accelerator neutrino data) respectively,
the measurements of mixing angle $\theta_{13}$ completed so far rely on neutrinos sourced from manmade facilities.
The T2K~\cite{T2K2011} and MINOS~\cite{MINOS2011}  Collaborations studied the appearance of electron neutrinos $\nu_e$ in a beam of muon neutrinos $\nu_\mu$
produced in particle accelerators, while the Double Chooz~\cite{DC2012}, Daya Bay~\cite{DB2012} and RENO~\cite{RENO2012} Collaborations
studied the disappearance of electron antineutrinos ${\bar \nu}_e$ produced from nuclear reactors in a nuclear power plant.
The measurement of $\theta_{13}$ is much more challenging than the other two mixing angles because $\theta_{13}$
is the smallest among the three  mixing angles. 
The limited flux of neutrino sources available from accelerators and reactors also constrained the measurement of $\theta_{13}$ to some extent.

The Daya Bay Collaboration first reported a successful measurement of mixing angle $\theta_{13}$ (with a statistical significance of $5.2 \sigma$) in 2012~\cite{DB2012} 
and the RENO Collaboration reported their results (with a statistical significance of $4.9 \sigma$) in the same year~\cite{RENO2012}.
More precise results from reactor neutrino experiments were reported in \cite{DBmore,RENOmore,DCmore}.
To compensate the limitation on the source of electron antineutrinos from nuclear power plants and to reduce systematic uncertainty 
in estimating absolute neutrino flux from nuclear power plants,
Daya Bay and RENO utilized the design of identical detectors placed in group of near and far away from the nuclear reactors.
There are 6 detectors in the first Daya Bay experiment~\cite{DB2012} and 2 in the RENO experiment~\cite{RENO2012}.
The number of nuclear reactors in both Daya Bay and RENO is 6.
More detectors have been placed in the late measurements~\cite{DBmore,RENOmore}.

The quantities measured in both experiments are number of electron antineutrinos at each detector over certain time period.
Thereby the numbers of data points in the first Daya Bay~\cite{DB2012} and RENO~\cite{RENO2012} experiments are 6 and 2, respectively.
While the use of multiple reactors and detectors improve the efficiency of the experiments, this will complicate the analysis of experimental data
since one has to deal with multiple sources of uncertainties which include
one for each reactor and two for each detector (one for the detector itself and one for the background of the detector).
Precise evaluation of uncertainties in these experiments is particularly important, considering the value of $\theta_{13}$ is very small.

The method used by the Daya Bay~\cite{DB2012} and RENO~\cite{RENO2012} Collaborations is know as pull method. This method is adopted 
with the intention to account for correlation of various systematic uncertainties.
The pull method has been used in studies that perform global fits of experimental data from different experiments in order to extract parameters
appearing in the fundamental theories, 
e.g. in the determination of parton distribution function of nucleon~\cite{Pull_PDF} 
and in the analysis of solar neutrino data from different experiments~\cite{Pull_other}.

The pull method introduces one pull factor for each systematic uncertainty. 
So there are 18 pull factors in Daya Bay and 10 in RENO experiments.
Together with the mixing angle $\theta_{13}$ and an overall normalization factor, there
are 20 and 18 parameters to be determined from the minimization of $\chi^2$ defined with 
6 and 2 points of data in the Daya Bay and RENO experiments respectively.
The Daya Bay and RENO Collaborations reported results on the mixing angle and the normalization factor while omitting any results on the pull factors.

It would be interesting to understand the role played by the pull factors. Are those pull factors all necessary?
What are the constraints on the values of pull factors? What information can those pull factors reveal about the fit? 
A number of issues were raised in \cite{KhanWR14} regarding the analysis of neutrino oscillation data from reactor neutrinos.
It was claimed \cite{KhanWR14} that they were unable to reproduce experimental groups' results due to their inability to access experimental details.
They reported an unexplained discrepancy between their results, obtained by following standard analysing procedures, 
and Daya Bay's and RENO's results.
They questioned the viability of the pull method used by the experimental groups in extracting parameters of interest when the number of data points is 
less than the number of parameters.
They constructed a $\chi^2$ in a similar way as the experimental groups, but involving fewer parameters than the ones used by the experimental groups, and 
demonstrated that minimization of this $\chi^2$ leads to multiple sets of parameters which do not always give consistent values for the parameters of interest.
The work reported in \cite{KhanWR14} casts some doubts on experimental groups' analyses of data.
The Daya Bay Collaboration has employed the covariance method of data analysis in their later studies \cite{DBmore}.

In this paper we address the issues with reproducibility of Daya Bay's results and the robustness of the pull method in analysing neutrino data.
We analyze the data following the same method described by the Daya Bay Collaboration \cite{DB2012} and 
present results for the mixing angle and the normalization factor, together with the values for the pull factors.
We investigate whether the extraction of parameters of interest are sensitive to the values of pull factors. 
Furthermore, we carry out an analysis with a $\chi^2$ defined with 5 fewer parameters than that used in Daya Bay's analysis 
after noticing the number of pull factors can be reduced due to identical detectors being used in the measurement.

We give a review of two approaches in defining $\chi^2$ with an emphasis on 
the constraints implied in the construction of $\chi^2$ in the pull approach in Section \ref{sec:review}.
An analysis following Daya Bay's approach is given in Section \ref{sec:DBfull},
while an analysis with a reduced $\chi^2$ is given in Section \ref{sec:DBreduced}. A summary is given in Section \ref{sec:summary}.

\section{The covariance approach vs pull approach}
\label{sec:review}

In this section we review two commonly used approaches in defining the $\chi^2$.
The underline assumption in defining a proper $\chi^2$ ({\it i.e.} it is a true measurement of goodness of fit) is
the defined $\chi^2$ must follow a normal (Gaussian) distribution.

For an experiment with $N$ observations (number of data points), correspondingly there are $N$ theoretical predictions.
The theoretical predictions normally depend on a set of parameters for the theory.
By comparing experimental measurements and theoretical calculations, one can test the theory and determine the parameters involved.
The $\chi^2$ is defined in terms of experimental observations $R_n^{\rm exp}$, theoretical predictions $R_n^{\rm the}$,
and uncorrelated and correlated uncertainties $u_n$ and $c_k$. The number of correlated uncertainties depends on the experiment setup.
$\chi^2$ is a numerical measurement of goodness of fit, {\it i.e.} the goodness of the theory describing the observations.

In the covariance approach, $\chi^2$ is defined as
\bea
\chi_{\rm cov}^2=\sum_{n,m=1}^N \left( R_n^{\rm exp} - R_n^{\rm the} \right)
\left[ \sigma_{nm}^2 \right]^{-1}
 \left( R_m^{\rm exp} - R_m^{\rm the} \right),
\label{eq:chi_cov}
\eea
where $\left[ \sigma_{nm}^2 \right]^{-1}$ is the inverse of $\sigma_{nm}^2$ --
a matrix of squared uncertainties constructed from uncorrelated and correlated uncertainties $u_n$ and $c_k$,
\bea
\sigma_{nm}^2=\delta_{nm} u_n u_m + \sum_{k=1}^K c_n^k c_m^k,
\label{eq:sigma}
\eea
with $K$ being the total number of correlated uncertainties.
The matrix of squared uncertainties is totally determined if all the uncertainties are well understood (given).
The parameters to be determined from the fit are the parameters appearing in the theory.
The number of degrees of freedom (d.o.f) is the number of observations (data points) minus the number of parameters.
An acceptable fit requires $\chi \sim {\rm d.o.f}$. 

It becomes increasingly challenging to invert the $N \times N$ matrix $\sigma_{nm}^2$ when the number of observations becomes large. 
The alternative pull approach introduces one parameter (``pull factor") for each correlated systematic uncertainty and defines $\chi^2$ as
\bea
\chi_{\rm pull}^2 &=& \sum_{n=1}^N \left( \frac{R_n^{\rm exp} - R_n^{\rm the} - \sum_{k=1}^K \xi_k c_n^k}{u_n } \right)^2
+\sum_{k=1}^K \xi_k^2 \nonumber \\
&\equiv& \chi_{\rm obs}^2+\chi_{\rm sys}^2.
\label{eq:chi_pull}
\eea

Despite the apparent difference between Eqs.~(\ref{eq:chi_cov}) and (\ref{eq:chi_pull}), it has been proven~\cite{Fogli2012} that the two approaches are 
equivalent, {\it i.e.} $\chi_{\rm cov}^2=\chi_{\rm pull}^2$. The key observation is that the minimization of Eq.~(\ref{eq:chi_pull}) with respect to $\xi_k$ 
leads to a set of $K$ linear equations for $\xi_k$ and the solutions of $\xi_k$ can be written in terms of a matrix which is related to the 
matrix $\sigma_{nm}$ given by Eq.~(\ref{eq:sigma}).
Thereby the pull factors introduced for $K$ systematic uncertainties are completely determined if all systematic uncertainties are given.

The pull approach has the advantage over the covariance approach in computation time when $K < N$ since the pull approach is equivalent to
the inversion of a $K \times K$ matrix while the covariance approach requires the inversion of a $N \times N$ matrix.
The pull approach is also applicable for the case $N<K$.
Another advantage of employing pull approach in data analysis is that it enables one to understand the effects of systematic uncertainties
in the measurement more clearly.
The effects from systematic uncertainty are expressed explicitly as the second term in Eq.~(\ref{eq:chi_pull}) ($\chi_{\rm sys}^2$)
while  the first term represents a measurement of difference between observations and theoretical predictions ($\chi_{\rm obs}^2$)

The parameters in Eq.~(\ref{eq:chi_pull}) include parameters needed for theoretical predictions which are of physics interest and 
pull factors. The number of pull factors introduced in Eq.~(\ref{eq:chi_pull}) is normally more than the number of parameters needed in 
theoretical calculations.
Although all pull factors are determined in principle when all uncertainties are known, 
in practice, the minimization of $\chi_{\rm pull}^2$ is carried out in the parameter space including parameter set for the theory and the pull factors.
When the number of data points is less than the number of parameters, as in the cases for the Daya Bay and RENO experiments,
the variational minimization usually will result in multiple sets of parameters that all give similar value for the $\chi^2$, {\it i.e.} having same level of goodness-of-fit.
One needs to rely on other information implied in the design of $\chi_{\rm pull}^2$ to distinguish these fits. 

Apart from goodness-of-fit test, there are other requirements for a (an) good/acceptable fit when using the pull method.
\begin{enumerate}
\item
All pull factors are small since the pull factors are Gaussian random variables with the expectations $\left< \xi_k \right>=0$ and $\left< \xi_k^2 \right>=1$.
\item
$\chi_{\rm sys}^2 << \chi_{\rm obs}^2$. A large value of $\chi_{\rm sys}^2$ means one or more systematic uncertainties dominates in achieving an agreement
between the data and theoretical calculations. 
\end{enumerate}
A violation of either condition usually indicates a great tension between experimental measurements and theoretical predictions.
We consider a fit that might pass the goodness-of-fit test ($\chi \sim {\rm d.o.f}$) but violates one or both above requirements is not acceptable. 

Even after applying the goodness-of-fit test and the above two criteria, there are still multiple acceptable fits depending on the minimization algorithm and/or initial guessed values 
for the parameters. This dependence should be insignificant for the quantities of interest.
To evaluate uncertainties associated with the existence of multiple acceptable fits, 
we adopt the following two methods:
1) We can perform a number of minimizations and calculate the statistical uncertainty in the standard way for multiple measurements.
2) An alternative way to avoid ambiguities in choosing the right fit from multiple sets of parameters is to treat $\chi_{\rm pull}^2$ as a function of parameters of interest.
For every point in the space of parameters of interest, a minimization is done with respect to all pull factors and
an acceptable minimization still need to meet all above requirements. In doing so, a plot of confidence level contours can be draw.

\section{Reanalysis of Daya Bay data}
\label{sec:DBfull}

The Daya Bay result for $\theta_{13}$ reported in \cite{DB2012} was obtained by minimizing a $\chi^2$ function, as was done by the other short-baseline neutrino oscillation experiments. 
The function used by the Daya Bay Collaboration [9] was,
\bea
\chi^2_{\rm DB}=\sum_{d=1}^6 \frac{\left[ M_d - T_d \left(1+\epsilon + \sum_{r=1}^6 \omega_r^d \alpha_r + \rho_d \right) +\eta_d  \right]^2}{M_d+B_d} 
+\sum_{r=1}^6 \frac{\alpha_r^2}{\sigma_r^2} + \sum_{d=1}^6 \left( \frac{\rho_d^2}{\sigma_d^2} + \frac{\eta_d^2}{\sigma_{db}^2}     \right),
\label{eq:DB}
\eea
where $M_d$ is the total number of antineutrinos detected by detector $d$,
$B_d$ is the measured background for detector $d$ and
$T_d$ is theoretical prediction for the number of antineutrinos reaching detector $d$ basing on neutrino oscillation theory
and thus depends on the mixing angle $\theta_{13}$.
$\epsilon$ is the normalization factor and $\omega_r^d$ is fractional contribution of antineutrinos detected by detector $d$ from reactor $r$.
$\sigma_r$, $\sigma_d$ and $\sigma_{db}$ are uncertainties associated with reactor $r$, detector $d$ and the background of detector $d$, respectively,
and $\alpha_r$,  $\rho_d$ and $\eta_d$ are the corresponding pull factors, respectively.

The theoretical prediction is simply the total flux at the detector multiplied by the survival probability.  The electron antineutrino survival probability is given by,
\bea
P_{\rm surv}(\bar{\nu}_{e}\rightarrow\bar{\nu}_{e})&=&1-\sin^2(2\theta_{12})\cos^4(\theta_{13})\sin^2\left(\frac{\Delta_{12}}{2}t\right) \nonumber \\
&& -\sin^2(2\theta_{13})\left[\cos^2(\theta_{12})\sin^2\left(\frac{\Delta_{13}}{2}t\right)+\sin^2(\theta_{12})\sin^2\left(\frac{\Delta_{12}}{2}t\right)\right],
\label{eq:Pfull}
\eea
where each $\theta_{ij}$ is a mixing angle, $t$ is the time of fly for $\bar{\nu}_{e}$ between the reactor and the detector,
 and $\Delta_{ij}$ is the difference in energy between mass states $i$ and $j$.  
Since detected neutrinos are always ultra-relativistic, the energy difference term is swapped out for the relativistic approximation,
and one has 
$\frac{\Delta_{ij}}{2}t\approx\frac{\Delta m^2_{ij}}{4E}L$ which depends on the mass-squared difference, the energy (which is assumed to be the same for all), and the baseline.

It should be noted that Eq.~(\ref{eq:Pfull}) is hierarchy dependent, therefore Daya Bay used the short baseline
approximation\footnote{The justification for this approximation is given in the supplemental material in~\cite{DB2012}.},
\bea
P=1-\sin^2(2\theta_{12})\cos^4(\theta_{13})\sin^2\left(\frac{\Delta m^2_{small}}{4E}L\right)-\sin^2(2\theta_{13})\sin^2\left(\frac{\Delta m^2_{big}}{4E}L\right),
\label{eq:PShort}
\eea
where the two mass squared differences are the measured values. 

The energy dependency of the probability was dealt with by performing a normalized integration of the probability over the nuclear emission and inverse beta decay absorption energy spectra,
which were obtained from~\cite{Espec} and~\cite{Cross-section} respectively.
(Not long after the publication of~\cite{DB2012}, Daya Bay found that the energy spectrum given in~\cite{Espec} was incorrect,
which prompted their use of the covariance method in later works~\cite{DBmore}.)

Equation~(\ref{eq:DB}) contains 20 free parameters of which $\theta_{13}$ and the overall normalization factor $\epsilon$ are of interest
while the other 18 pull factors are introduced to account for the correlation of the systematic uncertainties.
Daya Bay group reported only results for $\theta_{13}$ and $\epsilon$ but not for the pull factors.

We performed the fittings using the two methods described above: minimizing $\chi^2$ with respect to all parameters and minimizing $\chi^2$
with respect to the pull factors at fixed values of $\theta_{13}$ and $\epsilon$ and then finding the absolute minimum.
The two methods are equivalent as each is performing a global minimization; the first does so in one step while the second does so in two steps. 
The advantage of the first is that far fewer calculations are needed to arrive at ``the answer" while the second provides more information about
the behaviour of the function in the vicinity of the minimum.

The minimization algorithm used for each method naturally requires some initial  values for the parameters as inputs and is only capable of finding local minima. 
This was dealt with, in the case of the first method, by using psuedo-random inputs and taking the average of multiple minimizations,
assuming that the faux minima are uniformly distributed around the true minimum.
This can be confirmed if the value of $\chi^2$ produced when the obtained average values of the parameters are put into the function, is less than $\chi^2_{\rm ave}$.
The results can be improved by imposing some sensible conditions, such as requiring $\theta_{13}$ to be positive and the minima to fall within a certain range,
and choosing a reasonable range for the inputs. 

The fit described below used inputs taken from a uniform distribution ranging between 0 and 0.2 while minima were only accepted if they gave $\chi^2\leq5$
({\it i.e.} the values needed to be within the expected first confidence interval), and satisfied the two criteria discussed in Section \ref{sec:review} 
({\it i.e.}  all pull factors being small and $\chi_{\rm sys}^2 << \chi_{\rm obs}^2$).
We noticed that $\chi_{\rm sys}^2 << \chi_{\rm obs}^2$ is normally guaranteed when the other conditions for the acceptable fits are satisfied.
 A ``free" run  minimization without imposing any conditions discussed above could lead to output with abnormal large value of $\chi^2$ or large values for pull factors
with the associated result for the mixing angle being different from those reported by the Daya Bay Collaboration, a situation that has been reported in \cite{KhanWR14}.

An additional feature of using random inputs is that statistics can be calculated for each variable when multiple fits are performed,
which in turn will provide information about the topography of the function around the minimum.
The mean values and uncertainties corresponding to $95\%$ confidence interval for each variable obtained from 20 random input fits 
are presented in table~\ref{tablefull}.
The uncertainties listed in the table can be understood as the ``theoretical" uncertainties associated with the statistical method.
The uncertainty for the mixing angle is negligible small.
The values for the pull factors $\rho$ and $\alpha$ are on the order of $10^{-4}$,
while the values for the pull factor $\eta$ are on the order of of $10^{-1}$ but having uncertainties comparable with their mean values.
These results are consistent with the requirement that all pull factors should have an expectation value of zero.

\begin{table}
\caption{Results of random input fits with Eq.~(\ref{eq:DB})}
\centering
\begin{tabular}{l*{6}{c}}
Parameter& \multicolumn{2}{c}{$\chi^2$} &\multicolumn{2}{c}{$\sin^2(2\theta_{13})$} &\multicolumn{2}{c}{$\epsilon$$({\times}10^{-4})$} \\
\hline
Mean &\multicolumn{2}{c}{$3.872$} &\multicolumn{2}{c}{$0.08844$} &\multicolumn{2}{c}{$-2.3$}\\
90\%CI &\multicolumn{2}{c}{$0.073$} &\multicolumn{2}{c}{$0.00025$} &\multicolumn{2}{c}{2.7}\\
\hline
Parameter &$\rho_1$ &$\rho_2$ &$\rho_3$ &$\rho_4$&$\rho_5$ &$\rho_6$\\
\hline
Mean $({\times}10^{-4})$ &$4.90$ &$-6.45$ &$1.36$ &$4.573$ &$-0.24$ &$-4.335$\\
90\%CI $({\times}10^{-4})$ &$0.35$ &$0.07$ &$0.42$ &$0.039$ &$0.14$ &$0.031$\\
\hline
Parameter &$\alpha_1$ &$\alpha_2$ &$\alpha_3$ &$\alpha_4$ &$\alpha_5$ &$\alpha_6$ \\
\hline
Mean $({\times}10^{-4})$ &$-6.7$ &$-7.96$ &$7.8$ &$4.01$ &$8.4$ &$4.72$\\
90\%CI $({\times}10^{-4})$ &$4.9$ &$0.54$ &$4.6$ &$0.88$ &$5.5$ &$0.73$\\
\hline
Parameter &$\eta_1$ &$\eta_2$ &$\eta_3$ &$\eta_4$ &$\eta_5$ &$\eta_6$\\
\hline
Mean $({\times}10^{-1})$ &$2.0$ &$-1.2$ &$1.8$ &$4.2$ &$1.5$ &$4.5$\\
90\%CI $({\times}10^{-1})$ &$1.5$ &$2.4$ &$2.6$ &$1.6$ &$3.8$ &$2.1$\\ 
\end{tabular}
\label{tablefull}
\end{table}

Using the obtained values for the parameters as input for the function gives $\chi^2_{\rm min}=3.78\leq\chi^2_{\rm ave}=3.90$, $\chi_{\rm obs}^2=3.47$
and $\chi_{\rm sys}^2=0.31 <<\chi_{\rm obs}^2$. 
This satisfies two of the criteria for a valid goodness of fit test and indicates that the absolute minimum (or a close approximation) was obtained.
The pull factors are all quite small and their large variations relative to that of $\theta_{13}$ indicate that they are unimportant to the fit. 
This is further strengthened by the small value of $\chi_{\rm sys}^2$.
\begin{figure}
\includegraphics[width=\textwidth]{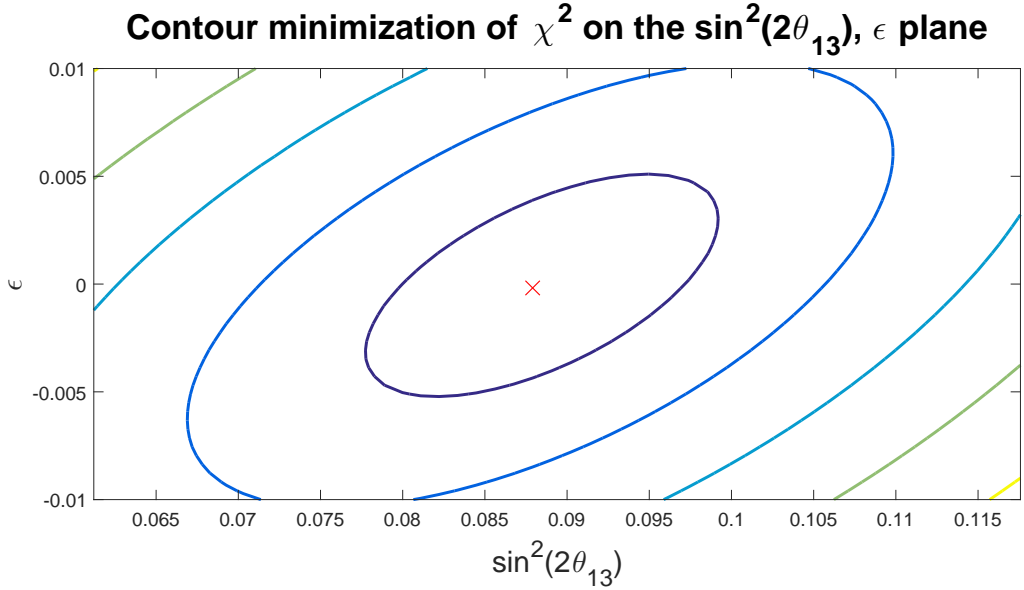}
\caption{(Color on line) Contour plot of the $\sin^2(2\theta_{13})$-$\epsilon$ plane for Eq.~(\ref{eq:DB}).  Each contour line represents $1\sigma$.  The red x is the minimum.}
\label{DBSCNPLT}
\end{figure}

For the second type of fit, minimization of $\chi^2$ for each point in the $\sin^2(2\theta_{13})$-$\epsilon$ space 
is carried out using the theoretical expectation value (zero) for the pull factors as the initial guess. 
A contour plot was produced in Fig.~\ref{DBSCNPLT} with each confidence interval being marked.
(For a two-variable fit, the $n$-th confidence interval is defined as $\sigma_n=\chi^2_{min}+n^2$.) 
The minimum found was $\chi^2=3.78$ at $\sin^2(2\theta_{13})=0.088_{-0.008}^{+0.009}$, and $\epsilon=0\pm0.004$, which is in agreement with the results obtained
with the first method we discussed above and in agreement with the results reported by the Daya Bay Collaboration.
A point of interest is that Fig.~\ref{DBSCNPLT} shows a greater variation along the $\epsilon$ axis than the $\sin^2(2\theta_{13})$ axis,
supporting the previous supposition in regards to the significance of the variability of the parameters in the random fit.  
Fig.~\ref{DBSCNPLT} also shows that the uncertainties of the two variables are correlated. 
A simple examination of Eq.~(\ref{eq:DB}) will reveal that the same effect will occur for all of the pull factors.

The antineutrino survival probability Eq.~(\ref{eq:Pfull}) is hierarchy dependent. We have checked that this hierarchy dependence has negligible effects on 
the extraction of the mixing angle $\theta_{13}$ and the normalization factor $\epsilon$.

\section{Daya Bay results with reduced parameters}
\label{sec:DBreduced}

We notice that the uncertainty of detector, $\sigma_d$, has the same values for all six detectors due to the fact that all six detectors are identical. 
Thus we argue that only one pull factor in association with the uncorrelated uncertainty of the detectors is required in the definition of $\chi^2$ function.
We define the reduced $\chi^2$ function as,
\bea
\chi^2_{\rm DB}=\sum_{d=1}^6 \frac{\left[ M_d - T_d \left(1+\epsilon + \sum_{r=1}^6 \omega_r^d \alpha_r + \rho \right) +\eta_d  \right]^2}{M_d+B_d} 
+\sum_{r=1}^6 \frac{\alpha_r^2}{\sigma_r^2} + \rho \frac{6}{\sigma_d^2} + \sum_{d=1}^6 \frac{\eta_d^2}{\sigma_{ab}^2}.
\label{eq:DBReduced}
\eea
The number of parameters needed in the fitting process is reduced from 20 to 15 by using Eq.~(\ref{eq:DBReduced}) in stead of Eq.~(\ref{eq:DB}).  Equation~(\ref{eq:DBReduced})
was minimized in the same two ways as Eq.~(\ref{eq:DB}).

\begin{table}
\caption{Results of random input fits with Eq.~(\ref{eq:DBReduced})}
\centering
\begin{tabular}{l*{6}{c}}
Parameter &$\chi^2$ &\multicolumn{3}{c}{$\sin^2(2\theta_{13})$} &$\epsilon({\times}10^{-5})$ &$\rho({\times}10^{-7})$\\
\hline
Mean &$4.1035$ &\multicolumn{3}{c}{$0.088551$} &$-1.93$ &$1.3$\\
90\%CI &$0.0086$ &\multicolumn{3}{c}{$0.000013$} &$0.61$ &$2.5$\\
\hline
Parameter &$\eta_1$ &$\eta_2$ &$\eta_3$ &$\eta_4$ &$\eta_5$ &$\eta_6$\\
\hline
Mean $({\times}10^{-4})$ &$-9.99$ &$-9.512$ &$5.484$ &$4.505$ &$4.233$ &$5.40$\\
90\%CI $({\times}10^{-4})$ &$0.12$ &$0.078$ &$0.041$ &$0.048$ &$0.061$ &$0.04$\\
\hline
Parameter  &$\alpha_1$ &$\alpha_2$ &$\alpha_3$ &$\alpha_4$ &$\alpha_5$ &$\alpha_6$\\
\hline
Mean $({\times}10^{-1})$ &$1.8$ &$2.6$ &$2.8$ &$4.1$ &$0.1$ &$2.3$\\
90\%CI $({\times}10^{-1})$ &$1.1$ &$1.5$ &$3.9$ &$2.7$ &$2.1$ &$1.4$\\
\end{tabular}
\label{tablerf}
\end{table}

The results from random input fits are given in Table~\ref{tablerf}.  Using the results as input for the function gives $\chi^2_{\rm min}=4.10=\chi^2_{\rm ave}$, $\chi_{\rm obs}^2=4.06$
and $\chi_{\rm sys}^2=0.05 <<\chi_{\rm obs}^2$ (some rounding occurred). 
Overall, the results are essentially the same as for Eq.~(\ref{eq:DB}).
\begin{figure}
\includegraphics[width=\textwidth]{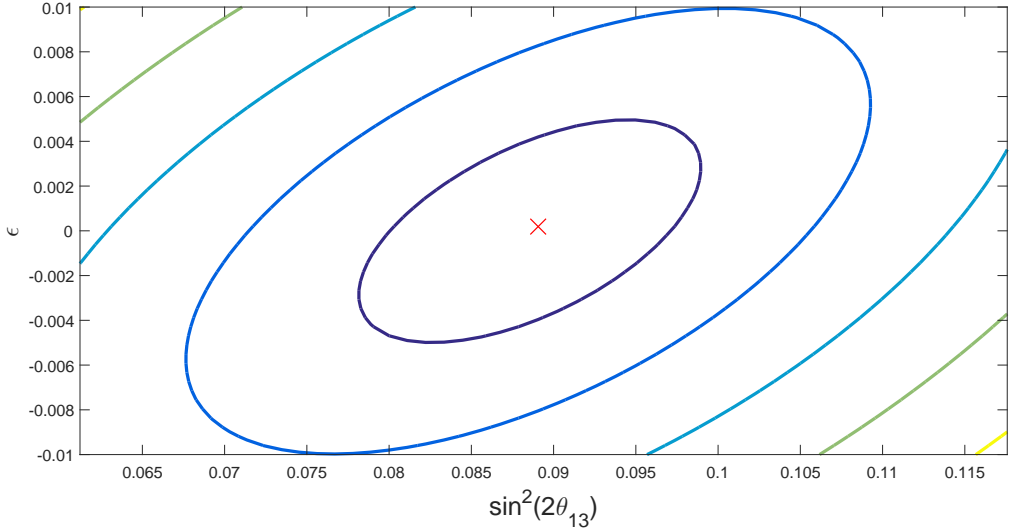}
\caption{(Color online) Contour plot of the $\sin^2(2\theta_{13})$-$\epsilon$ plane for Eq.~(\ref{eq:DBReduced}).  Each contour line represents $1\sigma$.  The red x is the minimum.}
\label{DBSCNRF}
\end{figure}
The second type of fit 
yields a contour plot as shown in Fig.~(\ref{DBSCNRF}).  The minimum found was $\chi^2=4.10$ at $\sin^2(2\theta_{13})=0.089^{+0.008}_{-0.009}$, $\epsilon=0\pm0.004$,
which is in agreement with the other method, as was the case with Eq.~(\ref{eq:DB}). 
Figure~\ref{DBSCNRF} shows the same $\sin^2(2\theta_{13})$-$\epsilon$ variation as Fig.~\ref{DBSCNPLT},
which is to be expected given the great similarity between Eqs.~(\ref{eq:DB}) and~(\ref{eq:DBReduced}).  Again, the results are essentially the same as for Eq.~(\ref{eq:DB}).
We conclude that Eq.~(\ref{eq:DBReduced}) is a suitable definition of $\chi^2$ in analysing Daya Bay's data.  

\section{Summary}
\label{sec:summary}

The successful measurement of mixing angle $\theta_{13}$ in the PMNS matrix is a significant progress in understanding neutrino oscillation phenomenon.
In order to eliminate uncertainty associated with neutrino fluxes from multiple nuclear reactors at a nuclear power plant, the experiment groups placed identical detectors
in different distance from the reactors, which unavoidably leads to the introduction of multiple sources of systematic uncertainties. 
The statistical method used in the data analyses by
the Daya Bay Collaboration and RENO Collaboration is based on variational minimization of a $\chi^2$ function defined
in terms of quantities of interest and pull factors which are introduced to deal with effects of systematic uncertainties.
There exist multiple sets of parameters from the minimization of the $\chi^2$ function. 

We reviewed the two statistical approaches in data analysis, the covariance approach and the pull approach, emphasizing the equivalence of these two approaches
and the other criteria needed for an acceptable fit apart from the normal goodness-of-fit test when the pull method is utilized.
Without imposing those criteria one is at the risk of extracting unreliable information for the quantity of interest.

We investigated the possible ambiguity in the extraction of mixing angle $\theta_{13}$ using the pull approach by performing an independent analysis of Daya Bay's data.
We reported results for  all parameters, including those of physics interest and pull factors.
The obtained results for the mixing angle and the normalization factor are in agreement with those reported by the Daya Bay Collaboration. 
Furthermore, we presented plots of confidence level contours in the space of the mixing angle and normalization factor.
Noticing some pull factors in the $\chi^2$ definition used by the Daya Bat Collaboration are surplus we presented results from fittings using a reduced $\chi^2$ function
which requires 5 fewer parameters. 

We have checked that the hierarchy dependence of the antineutrino survival probability
has negligible effects on the extraction of the mixing angle $\theta_{13}$ and the normalization factor~$\epsilon$.
A reanalysis of RENO's data using the methods presented in this paper will be given in future work.



\begin{thebibliography}{00}

\bibitem{DB2012}
	F.~P. An {\sl et al.}, Daya Bay Collaboration,
	\Journal{\PRL}{108}{171803}{2012}.

\bibitem{RENO2012}
	J.~K. Ahn {\sl et al.}, RENO Collaboration,
	\Journal{\PRL}{108}{191802}{2012}.

\bibitem{PMNS}
	See e.g., C. Patrignani et al. (Particle Data Group), 
	\Journal{\CPC}{40}{100001}{2016}.

\bibitem{T2K2011}
	K. Abe {\sl et al.}, T2K Collaboration,
	\Journal{\PRL}{107}{041801}{2011}.

\bibitem{MINOS2011}
	P. Adamson {\sl et al.}, MINOS Collaboration,
	\Journal{\PRL}{107}{181802}{2011}.

\bibitem{DC2012}
	Y. Abe {\sl et al.}, Double Chooz Collaboration,
	\Journal{\PRL}{108}{131801}{2012}.

\bibitem{DBmore}
	F.~P. An {\sl et al.}, Daya Bay Collaboration,
	\Journal{\CPC}{37}{011001}{2013}; {\sl ibid.}
	\Journal{\PRL}{112}{061801}{2014}; {\sl ibid.}
	\Journal{\PRL}{115}{111802}{2015}.

\bibitem{RENOmore}
	J.~H. Choi {\sl et al.}, RENO Collaboration,
	\Journal{\PRL}{116}{211801}{2015}.

\bibitem{DCmore}
	Y. Abe {\sl et al.}, Double Chooz Collaboration,
	\Journal{\JHEP}{1014}{086}{2014}.

\bibitem{Pull_PDF}
	D. Stump {\sl et al.},
	\Journal{\PRD}{65}{014012}{2001}.

\bibitem{Pull_other}
	See e.g.,
	S. Fukuda {\sl et al.}, The Super-Kamiokande Collaboration,
	\Journal{\PRL}{86}{5651}{2001};
	P. Huber, M. Lindner, T. Schwetz, and W. Winter,
	\Journal{\NPB}{665}{487}{2003}.

\bibitem{KhanWR14}
	A. N. Khan, D. W. McKay, and J. P. Ralston,
	\Journal{\IJMPA}{29}{1450109}{2014}.
	
\bibitem{Fogli2012}
	G. Fogli, E. Lisi, A. Marrone, D. Montanino, and A. Pallazo,
	\Journal{\PRD}{66}{053010}{2002}.


\bibitem{Espec}
	T. A. Mueller {\sl et al.},
	\Journal{\PRC}{83}{054615}{2011}. 

\bibitem{Cross-section}
	P. Vogel, and J. F. Beacom,
	\Journal{\PRD}{60}{053003}{1999}.

\end{thebibliography}
\end{document}